\documentclass[journal ]{new-aiaa}
\usepackage[utf8]{inputenc}
\usepackage{textcomp}

\usepackage{graphicx}
\usepackage{amsmath}
\usepackage[version=4]{mhchem}
\usepackage{siunitx}
\usepackage{longtable,tabularx}
\title{A high-performance lattice Boltzmann model for multicomponent turbulent jet simulations}

\author{Andrea Montessori \footnote{Associate Professor, Department of Civil, Computer Science and Aeronautical Technologies Engineering, andrea.montessori@uniroma3.it}}
\affil{Roma Tre University, Department of Civil, Computer Science and Aeronautical Technologies Engineering, Rome, 00146, Italy}
\author{Luiz A. Hegele Jr.}
\affil{Santa Catarina State University, Department of Petroleum Engineering, 88336-275, Balneário Camboriú, Brazil}
\author{Marco Lauricella}
\affil{Istituto per le Applicazioni del Calcolo CNR, via dei Taurini 19, 00185 Rome, Italy}
\begin{document}

\maketitle

\begin{abstract}
In this work an optimized multicomponent lattice Boltzmann (LB) model is deployed to simulate axisymmetric turbulent jets of a fluid evolving in a quiescent, immiscible environment over a wide range of dynamic regimes.
The implementation of the multicomponent lattice Boltzmann code achieves peak performances on graphic processing units with a significant reduction of the memory footprint,  
retains the algorithmic simplicity inherent to standard LB computing and being based on a high-order extension of the thread-safe lattice Boltzmann algorithm, it 
allows to perform stable simulations at vanishingly low viscosities.
The proposed approach opens attractive prospects for high-performance computing simulations of realistic turbulent flows with interfaces on GPU-based architectures.


\end{abstract}

\section*{Nomenclature}


{\renewcommand\arraystretch{1.0}
\noindent\begin{longtable*}{@{}l @{\quad=\quad} l@{}}
$f_i$  & lattice distribution \\
$\omega$ &    relaxation frequency \\
$\rho$ & fluid density \\
$u_\alpha$ & fluid velocity component\\
$w_i$ & lattice weight\\
$c_s$ & speed of sound\\
$\Pi_{\alpha\beta}$ & momentum flux tensor\\
$\sigma$ & surface tension\\
$\nu$ & kinematic viscosity\\
$\kappa$ & local curvature\\
$\phi$ & phase field\\
$Re$ & Reynolds number\\
$We$ & Weber number\\
$Oh$ & Ohnesorge number \\
$GLUPS$ & Billions of lattice update per second
\end{longtable*}}

\section{Introduction}

Understanding the dynamical features of turbulent jets is fundamental in many areas of engineering, including aeronautical, mechanical, and civil, to elucidate the underlying physical mechanisms present in a multitude of complex phenomena like combustion \cite{jiang2010physical}, propulsion\cite{oefelein2006large} and reactive flows\cite{grinstein1995three}.
In particular, breakup phenomena occurring in biphasic turbulent jets give rise to the onset of strikingly complex topological transitions \cite{eggers2018role,eggers1997nonlinear}, which impact on the performances of the systems in which turbulent jets occur.
Breakup processes are governed by a subtle competition among forces of different natures, namely, capillary, inertial, and drag forces and they encompass an extremely wide range of multi-scale physics phenomena, including the interaction of turbulence with interfaces \cite{constante2021direct}, capillarity \cite{montessori2019jetting,montessori2018elucidating}, and complex rheology induced by turbulent emulsification \cite{yi2023recent}  as well as heat transfer and phase change \cite{martinez2021new}.

Such an intrinsic multi-scale nature of the flow, together with the highly complex and rich physics exposed by the dynamical re-arrangements at the interface level, make difficult to unveil the plethora of different physical mechanisms occurring during the space-time evolution of the jet. Thus, elucidating the fundamental underlying physics of multicomponent jets inevitably asks for the use of accurate, high-fidelity, and efficient numerical models.

From the computational standpoint, the challenge is connected with the development of high-performance computing codes capable of i) spanning time and spatial scales significant to investigate the long-timescale behavior of the physics of fluids involved and ii) capturing the rich dynamical interactions among turbulence and fluid interfaces \cite{exasc2,exasc1}.

During the last couple of decades, the lattice Boltzmann method \cite{kruger2017lattice} (LB hereafter) has met with massive success in the computational fluid dynamics community as a hydrodynamic solver in kinetic disguise due to its conceptual and practical simplicity and its intrinsic efficiency \cite{succi,montessori2018lattice}. Indeed, at variance with standard hydrodynamic methods, in which the convective derivative is non-linear and non-local at a time, the LB splits non-linearity and non-locality between the two main operators, namely the collision (a local and non-linear operator) and the streaming step (linear and non-local and exact at machine precision). 

In this work, we extend a recent high-order, thread-safe version of the LB (TSLB)\cite{MONTESSORIjocs23, montessori2024high} ad-hoc optimized to run on shared memory architecture (typical of graphic processing units) by coupling the TSLB to  an interface capturing equation, solved via an efficient and shared-memory compliant LB solver. The multicomponent TSLB is employed here to simulate the evolution of an axisymmetric, multicomponent turbulent jet at $Re=5000$ and varying Weber numbers ranging from $10$ (dripping regime) to $2000$ (second-wind induced regime). 

\section{Methodology}

In this section, we briefly recall the main ingredients of the thread-safe LB framework. The TSLB is based on the cross-fertilization between a recently developed efficient LB strategy \cite{MONTESSORIjocs23} and a high-order variant of the regularized LB aimed at exploiting the recursivity of Hermite polynomials \cite{grad1949kinetic,grad1949note, malaspinas2015increasing} to reconstruct non-equilibrium moments up to third order. For an in-depth description of the method, the interested reader is referred to \cite{MONTESSORIjocs23,montessori2024high}
Before proceeding, a recap of the salient features of the lattice Boltzmann equation with the single-relaxation time approximation is given below (a thorough review of the LB equation can be found at Ref. \cite{kruger2017lattice, succi2018, montessori2018lattice}).

The discrete version of the Boltzmann equation reads:
\begin{equation} \label{lbe}
    f_i(x_\alpha+c_{i\alpha}\Delta t,t + \Delta t)=f_i(x_\alpha,t) + \omega(f^{eq}_i(\rho,\rho u_\alpha)-f_i(x_\alpha,t))
\end{equation}
where $x_\alpha$ and $t$ denote lattice position and time step, respectively, $f_i(x_\alpha,t)$ is a set of distribution functions, $i=1,..,q$ is an index spanning the $q$ discrete velocity vectors $c_{i\alpha}$ of the lattice and $f_i^{eq}(\rho,\rho u_\alpha)$ is a set of discrete thermodynamic equilibria obtained via a second-order expansion in Mach number of the continuous Maxwell-Boltzmann distribution. Also, $\rho$ is the fluid density, and $\rho u_{\alpha}$ is the fluid momentum. 
The equilibrium distribution functions, $f_i^{eq}$, is given by \cite{kruger2017lattice}
\begin{equation}\label{lbequil}
    f_i^{eq}=w_i \rho\left(1 + \frac{c_{i\alpha} u_\alpha}{c_s^2} + \frac{(c_{i\alpha} c_{i\beta} - c_s^2\delta_{\alpha\beta})u_\alpha u_\beta}{2c_s^4}   \right )
\end{equation}
where $w_i$ is a normalized set of weights and $c_s^2=1/3$ is the sound speed of the model. As can be seen from eq. (\ref{lbe}), the standard LB algorithm can be split into two main steps, namely a relaxation towards a local Maxwell-Boltzmann equilibrium occurring at a characteristic rate $\omega$ (right-hand side of eq.(\ref{lbe}), local and non-linear operation) and a free streaming of the discrete distributions along linear characteristics (linear and non-local and exact at machine precision).
The hydrodynamic fields of interest, i.e., density $\rho$, linear momentum $\rho u_\alpha$ and momentum flux tensor $\Pi_{\alpha\beta}$, can be retrieved by computing the statistical moments of the distribution functions $f_i$ \cite{succi2018}. 

To note that a multi-scale expansion in the smallness parameter (i.e., the Knudsen number )\cite{chapman1990mathematical} can be performed to prove that the lattice Boltzmann equation recovers a set of partial differential equations governing the conservation of mass and momentum of a volume of fluid, in the limit of weak compressibility (small Mach numbers): 

\begin{equation}
    \partial_t \rho +  \partial_{\alpha}\rho u_\alpha=0,
\end{equation}
\begin{equation}
    \frac{\partial \rho u_\alpha}{\partial t} + \partial_{\beta} (\rho u_\alpha u_\beta) = \partial_{\alpha}(-p \delta_{\alpha\beta} + \rho\nu(\partial_{\beta} u_\alpha + \partial_{\alpha} u_\beta))
\end{equation}
where $p=\rho c_s^2$ is the pressure, and $\nu=c_s^2(1/\omega-0.5)$ is the kinematic viscosity. 


Equation \ref{lbe} suggests that, in a lattice BGK relaxation process, any post-collision distribution can be expressed as a weighted sum of the equilibrium and non-equilibrium parts of the $i^{th}$ probability distribution function (pdf), and by exploiting the following decomposition of the discrete pdf, $f_i=f_i^{eq} + f_i^{neq}$, one can write \cite{MONTESSORIjocs23}:
\begin{equation}\label{LBexp}
    f^{post}_{i} =f^{pre}_{i} + \omega(f^{eq}_{i} - f^{pre}_{i})=f^{eq}_{i} + (1-\omega)f^{neq}_{i}
\end{equation}

A direct consequence of Eq.\ref{LBexp} is the possibility to reconstruct the set of distribution functions by decoupling pre- ($f^{pre}_{i}$) and post-collisional ($f^{post}_{i}$) states, with evident advantages for the implementation of highly efficient LB models on shared-memory architectures (SMA) present in graphic processing units (GPUs) \cite{MONTESSORIjocs23,montessori2024high}.
Indeed, Eq.(\ref{LBexp}) eliminates data dependencies arising during non-local read and write operations, avoiding race condition problems.

It is worth noting that, usually, to avoid race conditions in a LB algorithm an A-B strategy can be employed at the expense of  doubling the allocated memory of LB simulations on SMA. A possible path to reduce the memory footprint of LB models is to resort to more exotic streaming strategies \cite{bailey2009accelerating, geier2017esoteric, lehmann2022esoteric},  at the expense of the coding simplicity and readability of the LB code.
On the other hand, Eq. (\ref{LBexp}) can be straightforwardly extended to include the streaming step as follows:

\begin{equation}\label{pushLB}
    f_{i}(x_\alpha+c_{i\alpha},t+1) = f^{eq}_{i}(x_\alpha,t) + (1-\omega)f^{neq}_{i}(x_\alpha,t)
\end{equation}

The second pivotal step to transform the algorithm in Eq. \ref{pushLB} into a thread-safe operation on SMA is to reconstruct the non-equilibrium distributions by projecting them onto a suitable set of Hermite basis, $f^{neq}_i=w_i \sum_n \frac{1}{c_s^{2n} n!}a_{neq,\alpha 1...\alpha n}^n \mathcal{H}_{i,\alpha1...\alpha n}^n$ where $\mathcal{H}^n$ is the n-th order Hermite basis and $a_{neq,\alpha1...\alpha n}^n=\sum_i f^{neq}_i \mathcal{H}_{i,\alpha1 ... \alpha n}^n$ the corresponding Hermite expansion coefficients. Both $a^n$ and  $\mathcal{H}^n$ are rank $n$ tensors.

Thus, the set of non-equilibrium distributions can be compactly written as:

\begin{equation}
    f^{neq}_i=\frac{w_i}{2 c_s^4} (c_{i\alpha}c_{i\beta} - c_s^2\delta_{\alpha\beta}) \Pi^{neq}_{\alpha\beta}
\end{equation}

The above relation permits to rewrite the lattice dynamics reported in Eq.(\ref{LBexp}) by storing, at each time step and for each lattice node, three macroscopic quantities, namely, one scalar ($\rho$), three vector components ($\rho \mathbf{u}$) and six components of a symmetric, tensor of rank two ($\Pi^{neq}_{\alpha\beta}$). Furthermore, note that density and velocity fields are usually allocated in the standard LB approach to compute the equilibrium distribution at each time step. Thus, the extra memory requirement in our approach reduces to six extra arrays needed to store the components of the second order symmetric tensor $\Pi^{neq}_{\alpha\beta}$. Thus, in order to perform a lattice Boltzmann simulation on a D3Q27 lattice (single component fluid) we need to allocate $36$ three-dimensional arrays instead of $63$ (A-B strategy). 


Finally, the full stream and collision step (fused step) can be written as follows:

\begin{multline} \label{fullstrcoll}
    f_{i}(x_\alpha+c_{i\alpha},t+1) = \rho w_i [ 1 + \frac{c_{i\alpha}\cdot u_\alpha}{c_s^2} + \frac{(c_{i\alpha}c_{i\beta}-c_s^2\delta_{\alpha\beta}) u_\alpha u_\beta}{2 c_s^4}] +  (1 - \omega)\frac{w_i}{2 c_s^4} (c_{i\alpha}c_{i\beta} - c_s^2\delta_{\alpha\beta}) \Pi^{neq}_{\alpha\beta}
\end{multline}

\begin{figure}
    \centering
    \includegraphics[scale=0.8]{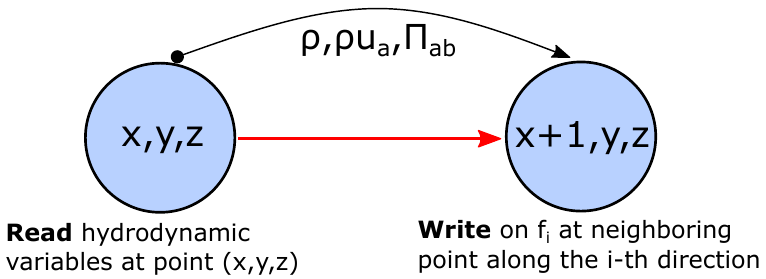}
    \caption{Sketch of the thread-safe LB update algorithm. Rather than explicitly streaming distributions along the lattice directions, the algorithm reconstructs the post-streamed distribution using the neighboring macroscopic hydrodynamic fields, which are employed to rebuild the full set of distributions.}
    \label{fig:threadsafe}
\end{figure}

In other words, the thread-safe LB, rather than explicitly streaming distributions along the lattice directions, aims at reconstructing the post-streamed/post-collided distribution by reading the neighboring macroscopic hydrodynamic fields, which are then employed to rebuild the complete set of distributions at the neighboring lattice node (see Fig.\ref{fig:threadsafe}).

\begin{figure}
    \centering
    \includegraphics[scale=0.5]{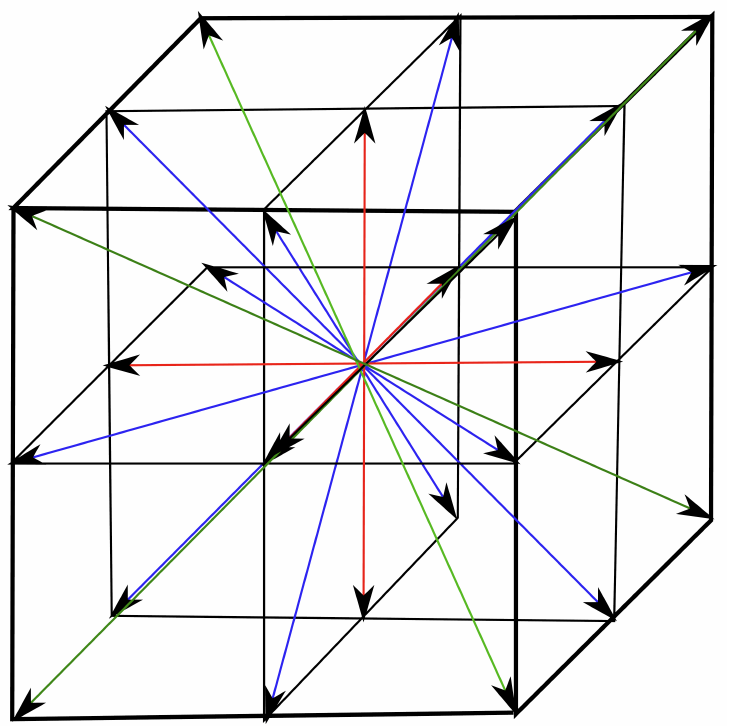}
    \caption{Graphical representation of the D3Q27 lattice. Red arrows stand for the velocity group having magnitude $|c|=\sqrt{c_x^2+c_y^2+c_z^2}=1$, the blue one $|c|=\sqrt{2}$ and the green arrows $|c|=\sqrt{3}$ }
    \label{fig:d3q27}
\end{figure}

It is interesting to note that, in the framework of the thread-safe LB, it is possible to implement Dirichlet and Neumann-like boundary conditions by exploiting the concept of non-equilibrium extrapolation and Grad's reconstruction of missing data at the boundary \cite{montessori2024high, chikatamarla2006grad,guo2002extrapolation} with no complication coding-wise.
Indeed, the missing distribution at the boundary can be obtained via the following relation:

\begin{equation} \label{original_neq}
    f_i(x_{B\alpha},t)=f_i^{eq}(x_{B\alpha},t)+(1-\omega)f_i^{neq}(x_{F\alpha},\tau)
\end{equation}

where $f_i^{neq}(x_{F\alpha},t)$ is the i-th non-equilibrium distribution in the neighboring fluid node to the boundary, which can be reconstructed by extrapolating non-equilibrium moments of the distribution from the neighboring fluid nodes along the lattice directions:

\begin{equation} \label{neq_final}
    f_i^{neq}(x_{F\alpha},t)=w_i\sum_n \frac{1}{c_s^{2n} n!}\mathcal{H}_{i\alpha 1...\alpha n}^n a_{neq,\alpha 1...\alpha n}^n(x_{F\alpha},t)
\end{equation}

while the equilibrium part is evaluated from the macroscopic density (pressure) and/or linear momentum values imposed at the boundary nodes. 
An important advantage of such an  
approach lies in the possibility of imposing boundary conditions directly on 
pressure and momentum, as in standard Navier-Stokes solvers, while approximating the momentum flux tensor (containing information on the deviatoric part of the stress tensor) with its value in the nearest-neighbor bulk node. 

Operationally, one has to impose values of the macroscopic low-order moments at the boundary to compute the equilibrium distribution while the non-equilibrium is readily extrapolated from the bulk. Then, equation \ref{pushLB} is employed to stream back the populations in bulk.


\subsection{\label{sec:recreg} High-order regularized lattice Boltzmann based on recursivity of Hermite polynomials}

As shown in the previous section, the distribution can be split into its equilibrium and non-equilibrium parts, which can be represented through Hermite polynomials:

\begin{equation}
    f_i^{eq}=w_i \sum_n \frac{1}{c_s^{2n} n!}a_{eq,\alpha 1...\alpha n}^n \mathcal{H}_{i\alpha1...\alpha n}^n
\end{equation}
\begin{equation}
    f_i^{neq}=w_i \sum_n \frac{1}{c_s^{2n} n!}a_{neq,\alpha 1...\alpha n}^n \mathcal{H}_{i\alpha1...\alpha n}^n
\end{equation}

The equilibrium and non-equilibrium Hermite coefficients are defined as:

\begin{equation} \label{hermite_eq}
   {a}_{eq,\alpha 1,...,\alpha n}^n=\sum_i \mathcal{H}_{i\alpha 1,...,\alpha n}^n f_i^{eq}
\end{equation}
and 
\begin{equation}
    a_{neq,\alpha 1,...,\alpha n}^n=\sum_i \mathcal{H}_{i\alpha 1,...,\alpha n}^n f_i^{neq}
\end{equation}

As far as the equilibrium Hermite coefficients are concerned, one can construct the $n-th$ order coefficient by exploiting the following relation \cite{shan2006kinetic}:

\begin{equation}
    {a}_{eq,\alpha 1,...,\alpha n}^n={a}_{eq,\alpha 1,...,\alpha n-1}^{n-1}u_{\alpha n}
\end{equation}

being ${a}_{eq}^0=\rho$, $a_{eq}^1=\rho u_\alpha$ and $a_{eq}^2=\rho u_\alpha u_\beta + \rho c_s^2 \delta_{\alpha\beta}$ .
Since the Hermitian D3Q27 correctly recovers macroscopic moments up to order three ($n=3$)\cite{succi, kruger2017lattice}, the use of eq.(\ref{hermite_eq}) expanded up to $n=3$ \cite{shan2006kinetic}, permits to define the equilibrium functions as:

\begin{multline}\label{eqts}
    f_i^{eq}=\rho w_i [ 1 + \frac{c_{i\alpha}\cdot u_\alpha}{c_s^2} + \frac{(c_{i\alpha}c_{i\beta}-c_s^2\delta_{\alpha\beta}) u_\alpha u_\beta}{2 c_s^4} \\ 
    + \frac{(c_{i\alpha}c_{i\beta}c_{i\gamma}-c_{i\gamma}c_s^2\delta_{\alpha\beta} - c_{i\alpha}c_s^2\delta_{\beta\gamma} - c_{i\beta}c_s^2\delta_{\alpha\gamma}) u_\alpha u_\beta u_\gamma}{6 c_s^6}]
\end{multline}

The above third-order, discrete Maxwellians allow the recovery of hydrodynamic moments up to order $3$, i.e.,

\begin{equation}
    Q_{\alpha\beta\beta}=\sum_i f_i^{eq}c_{i\alpha}c_{i\beta}c_{i\beta}
\end{equation}

namely, a third-order tensor representing a flux of momentum flux \cite{montessori2015lattice}. Despite such a higher-order moment having no counterpart in the Navier-Stokes equation, the use of extended equilibria increases both the stability and accuracy of an LB solver \cite{montessori2024high, malaspinas2015increasing, montessori2015lattice}.

As shown in \cite{grad1949note, shan2006kinetic}, it is possible to use recursivity of Hermite polynomials such that the Hermite coefficients (of order larger than $2$) can be written as:

\begin{equation}
    a^n_{{neq},\alpha 1,...\alpha n}= a^{n-1}_{{neq},\alpha 1,...\alpha n-1} u_{\alpha n} + (u_{\alpha 1} \cdot\cdot\cdot u_{\alpha n-2} a^{2}_{{neq},\alpha n-1 \alpha n})
\end{equation}

Recalling that $a_{{neq},\alpha\beta}^{2}=-\frac{1}{\omega c_s^2} (\partial_{\alpha}\rho u_\beta+\partial_{\beta}\rho u_\alpha))$, the non-equilibrium set of distributions can be explicitly retrieved by incorporating non-equilibrium hydrodynamic moments up to the third-order:

\begin{multline} \label{noneqts}
    f_i^{neq}=\frac{c_{i\alpha}c_{i\beta} a^2_{neq,\alpha\beta}}{2 c_s^4} + \\ \frac{(c_{i\alpha}c_{i\beta}c_{i\gamma}-c_{i\gamma}c_s^2\delta_{\alpha\beta} - c_{i\alpha}c_s^2\delta_{\beta\gamma} - c_{i\beta}c_s^2\delta_{\alpha\gamma}) (a^2_{{neq},\alpha\beta} u_\gamma + u_\alpha a^2_{{neq},\beta\gamma} + u_\beta a^2_{{neq},\alpha\gamma} )}{6 c_s^6}
\end{multline}


\subsection{A lattice Boltzmann implementation of the interface capturing equation}

The last ingredient needed to capture the evolution of a fluid-fluid interface and, in turn, to code for the build-up of a surface tension, compliant with the Young-Laplace law, is an interface capturing scheme. 
First of all, it is worth noting that the term in the NSE responsible for the onset of a positive surface tension among two immiscible fluids is represented by the following divergence operator:

\begin{equation}
    F_\alpha^\sigma=-\sigma(\nabla\cdot n_\alpha) \delta_{int}
\end{equation}

where $n_\alpha$ is the local normal to the interface. Once a suitable phase field is defined, the $n_\alpha$ is defined via the following relation:

\begin{equation}
    n_\alpha=\nabla\cdot\frac{\nabla \phi}{|\nabla \phi|}
\end{equation}

being $\phi$ the phase field.
To capture the dynamics of the phase field, we need to define a suitable evolution equation for $\phi$, which is supposed to be governed by the following advection diffusion anti-diffusion equation:

\begin{equation} \label{diff_antidiff}
    \partial_t \phi  + u_\alpha \partial_{x_\alpha} \phi=D_{int}\partial_\alpha \partial_\alpha \phi + R \delta_s
\end{equation}

where $R\delta_s=\gamma \nabla\cdot [\phi(1-\phi) n_\alpha]$ is a source term acting along the normal to counteract interfacial diffusion, as in \cite{OLSSON2007785,reis2022lattice}, $\gamma$ a proportionality (anti-diffusion) constant and $D_{int}$ is the diffusivity of the interface. 

Without losing generality, equation (\ref{diff_antidiff}) can be efficiently solved via a lattice Boltzmann method on a $D3Q7$ lattice, equipped with first-order truncated equilibria and by relaxing the non-equilibrium part of the distribution with a unitary relaxation time.
The resulting LB equation reads as follows:

\begin{equation}\label{intcapturing}
    g(x_\alpha+c_{\alpha i},t+1)=g^{eq}(\phi,u_\alpha) + \Lambda_i
\end{equation}

where $g(x_\alpha,t)$ is the distribution, whose zeroth order moment is the local phase-field, and $\Lambda_i$ is the $i-th$ component of the anti-diffusion term needed to maintain the interface between the two fluids sharp and is defined as:

\begin{equation} \label{antidiff}
    \Lambda_i=w_i \gamma \phi(1-\phi) c_{i \alpha }n_\alpha
\end{equation}

Thus, eq.(\ref{intcapturing}) exploits the concept underlying the thread-safe strategy, namely, reading from hydrodynamics fields and writing in pdf arrays, thus allowing for an efficient and safe implementation of interface capturing dynamics via LB on shared memory architectures, again avoiding race condition problems.

To note, in this work, we simulated the evolution of a turbulent jet of fluid, labeled A, within a quiescent environment, B. The two immiscible interacting fluids share the same viscosity and density. We are currently extending the present model to code for variable density and viscosity ratios and we will present such extended model in a future work.

\section{Results}

The multicomponent TSLB is employed to simulate the space-time evolution of a two-component axisymmetric turbulent jet featuring an inlet Reynolds number $Re_{jet}=\frac{U_{jet}D}{\nu}=5000$, being $U_{jet}$ the inlet velocity, $D=40$ (diameter of the round nozzle, in grid units) and $\nu$ the kinematic viscosity of the fluid. The inlet Weber number has been varied systematically, from $We=10$ to $We=2000$, to assess the capability of the proposed model to reproduce the main breakup regimes occurring in turbulent jets of immiscible fluids at different Ohnesorge (i.e., $Oh=\sqrt{We}/Re$) numbers.

\begin{figure}
    \centering
    \includegraphics[scale=0.8]{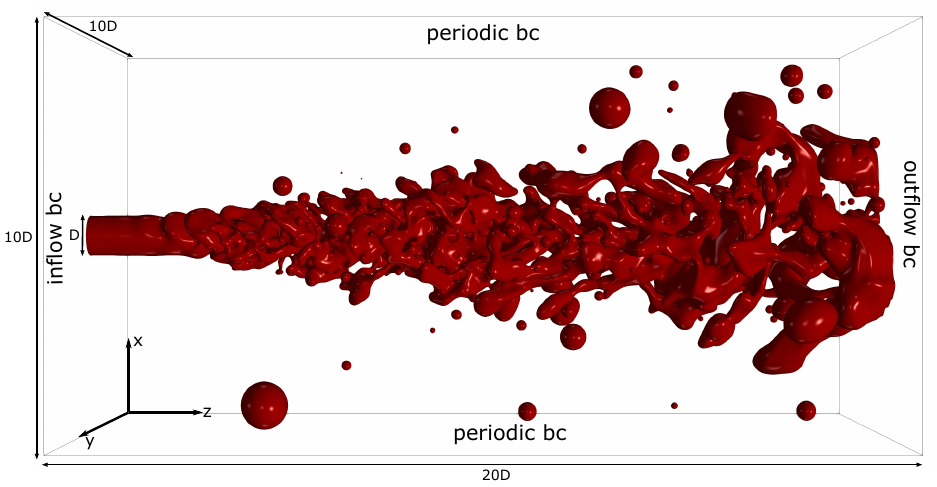}
    \caption{Snapshot of a turbulent jet simulation performed at $Re=5000$ and $We=500$. The dimensions of the domain are scaled by the length of the diameter of the nozzle, $D=40$ grid points. The domain is composed of $10D\times 10D \times 20 D= 128\times10^6$ grid points. }
    \label{fig:domain}
\end{figure}

\begin{figure}
    \centering
    \includegraphics[scale=0.6]{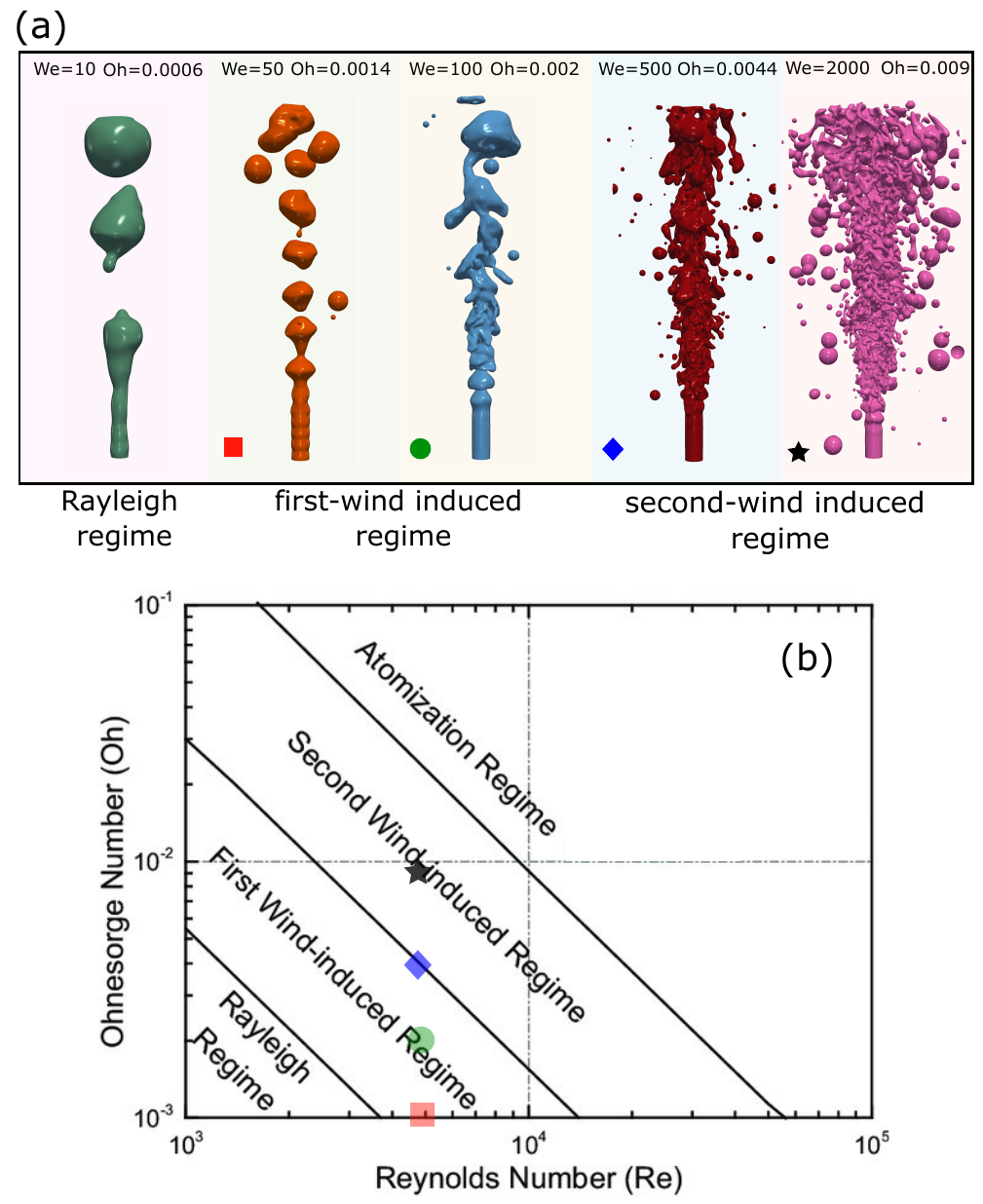}
    \caption{Breakup regimes simulated via the multicomponent, high-order TSLB model, from Capillary induced (turbulent Rayleigh) regime to second-wind induced regime. The Reynolds number has been kept fixed while the Weber has been varied by changing both the inlet velocity and the surface tension in the model. (b) The plot reports the Ohnesorge-Reynolds map for different types of breakup regimes, and the symbols denote the position of the four simulations in the Oh-Re plot (The Rayleigh regime is not reported since it falls outside the Oh-Re map)}
    \label{fig:regimes}
\end{figure}

To this aim, the plot of figure \ref{fig:regimes}(b) reports the $Oh$ versus $Re$ plot, which is employed to characterize the dynamical regime of turbulent jets at varying Reynolds and Weber numbers. Indeed, depending on the inertial forces, surface tension, and aerodynamic (viscous drag) forces acting on the jet, several regimes can be identified, namely the Rayleigh regime, or drip flow regime, the first wind-induced and second wind-induced regime, arriving at the atomization regime. In this work, Rayleigh to second-wind-induced regimes have been simulated.
Before proceeding with the discussion of the main results, we note that, throughout all the simulations, the convective Mach number, defined as $Ma=U_{jet}/c_s$, has always been kept well below the threshold value $Ma_c=0.6$, above which compressibility effects may not be negligible, affecting the overall dynamic of the jet \cite{wang2010direct}.

\subsubsection*{Breakup Regimes}
As shown in Fig. \ref{fig:regimes}(a), the multicomponent TSLB is capable of reproducing different breakup regimes, in agreement with the Oh-Re map \cite{ohnesorge2019formation,trettel2020reevaluating}. For clarity, in table \ref{tab:table}, we report the conditions employed in each simulation.

\begin{table}
    \centering
    \begin{tabular}{|c|c|c|c|c|c|c|c|}
    \hline 
         $sim$ & $U_{jet}$ & $\sigma$ & $\nu$ & $D$  & $nx\times ny \times nz$  & $Re$ & $We$ \\
         1&  0.01    & $4\cdot10^{-4}$ & $6.7\cdot10^{-5}$ & 40 & $400\times400\times800$ & 5000 & 10 \\
         2&  0.0267  & $3.2\cdot10^{-4}$ & $1.67\cdot10^{-4}$ & 40 & $400\times400\times800$ & 5000 & 50 \\
         3&  0.0267  & $2\cdot10^{-4}$ & $1.67\cdot10^{-4}$ & 40 & $400\times400\times800$ & 5000 & 100\\
         4& 0.05     & $2\cdot10^{-4}$ & $4\cdot10^{-4}$ & 40 & $400\times400\times800$ & 5000 & 500 \\
         5& 0.05     & $5\cdot10^{-5}$ & $4\cdot10^{-4}$ & 40 & $400\times400\times800$ & 5000 & 2000 \\
         \hline 
    \end{tabular}
    \caption{Simulations data. Physical paramters are expressed in lattice units.}
    \label{tab:table}
\end{table}

A quick glance at fig. \ref{fig:regimes}(a) suggests how, at sufficiently low Weber numbers, a dripping regime can be observed, even at  Reynolds numbers high enough to sustain a turbulent velocity field. This regime emerges from a subtle competition between surface tension and liquid inertia with negligible influence of the drag forces. In the dripping regime, at the breakup, the characteristic size of the newly formed droplet is generally equal, or slightly larger, than the nozzle diameter, as can be observed from the leftmost snapshot in fig. \ref{fig:regimes}(a).

As the Weber number increases, the jet passes into the second region of the Oh-Re map, i.e., the first-wind-induced regime. In this case, viscous effects are no longer negligible, and the shearing actions at the interface enhance wave disturbances, contributing to the breakup of the liquid thread. In passing from $We=100$ (sim 2) to $We=500$ (sim 3) we observe a transition between a "laminar" and "turbulent" first-wind induced regime \cite{trettel2020reevaluating} in which the fluid-fluid interface appears to be more wavy and jagged due to the onset of Kelvin-Helmoltz instabilities downstream the inlet nozzle within the potential core region. Such instabilities are triggered by the increasingly dominant action of viscous drag and inertia over surface tension forces.  Nonetheless, the capillary forces are still strong enough to prevent the complete disintegration of the liquid jet. Indeed, the jet exiting from the nozzle appears to be still radially confined, with the maximum spreading still limited by the size of the nozzle. 

Upon further increasing the Weber number, the jet transits to the second-wind-induced regime, characterized by a more evident atomization of the fluid downstream of the potential core. In this regime, the elongation and deformation of the detached droplets increase as the Weber increases, inducing secondary breakup events, which result in a reduction of the size of the newly formed drops as the Weber increases. Such a regime is characterized by a spreading of smaller droplets in the crossflow directions ($x-y$ plane) due to the radial motion induced by turbulent diffusion.

\subsubsection*{Self-similarity in multicompoenent turbulent jets}

\begin{figure}
    \centering
    \includegraphics[scale=0.8]{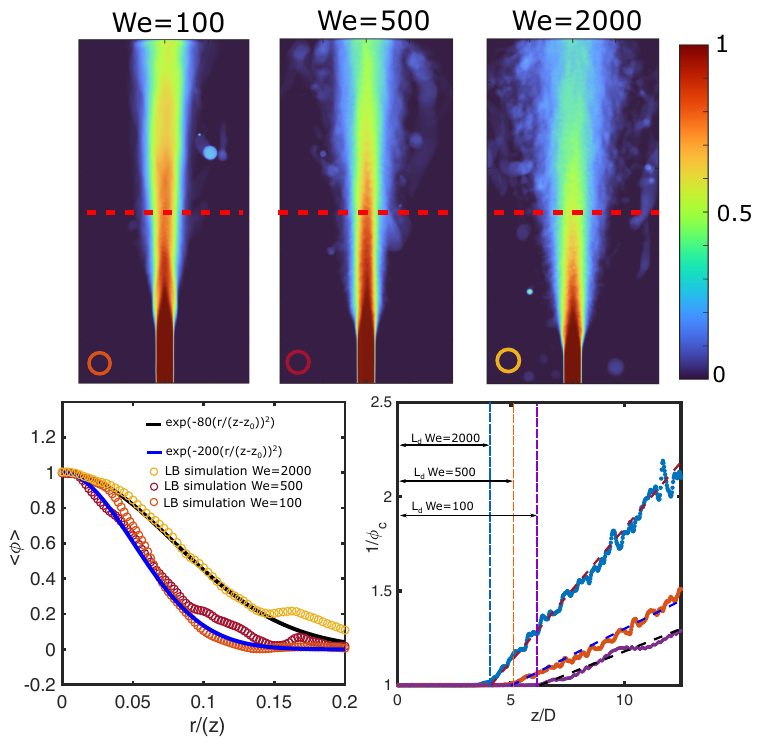}
    \caption{Shadowgraph images of the time-average jets for three different Weber numbers. The plots below report the time-average profiles of the phase field taken along the dashed lines reported in the shadowgraphs along with two Gaussian distributions $k_{int}=80$ (\cite{papanicolaou1988investigations}) and $k_{int}=200$ (left plot). The right plot reports the inverse of the centerline phase field ($\/\phi_c$). The numerical results have been fitted with straight lines of different slopes. The net effect of the larger magnitude of the capillary forces at lower Weber numbers is to counteract viscous drag and inertia, namely the dynamical actions responsible for the jet breakup and subsequent atomization at larger We, and, consequently, to increase the length of the potential core region (distance from the nozzle at which interfacial instabilities take over and jet breaks.) }
    \label{fig:shadowgraphs}
\end{figure}

Figure \ref{fig:shadowgraphs} shows the shadowgraphs of the time-averaged phase field, taken on the mid $x-z$ plane, for three different Weber numbers. Their inspection allows us to gain a deeper understanding of how the competition of capillary, viscous, and inertial forces shapes the dynamical characteristics of the jet \cite{ibarra2020near}.
In particular, we can observe that, at sufficiently low Weber numbers ($We=100$), the fluid exits the nozzle with a limited lateral diffusion while, as the Weber increases, turbulence takes over, favoring a fine emulsion of the jet together with a more evident radial growth, as occurs in momentum jets.

The above observations can be put on a more quantitative ground by plotting the time-averaged phase field, $<\phi>$, along the radial direction at a given distance from the nozzle. As shown in the leftmost plot of figure \ref{fig:shadowgraphs}, the three profiles ($We=100-2000$) are well-fitted by Gaussian curves with different standard deviations. In particular, the transition from the first to the second wind induces a wider jet spreading as clearly seen from the left plot. Interesting to observe, the best fit of $<\phi>$ at $We=2000$  is obtained with a characteristic exponent $K=80$,  in agreement with the experimental findings reported in \cite{papanicolaou1988investigations}. Thus, as the Weber number increases, i.e., as the relative magnitude of surface tension forces with respect to inertial forces decreases, the multicomponent jet behaves as a turbulent buoyant plume, and the average dynamics of the atomized droplets could be, in principle, described via a coarse-grained turbulent advection-dispersion-like equation. A second observation is that the potential core region increases as the Weber number of the jet decreases. This can be assessed by inspecting the right plot in fig \ref{fig:shadowgraphs}, in which the vertical dashed lines denote the length (distance from the inlet nozzle) of the potential core region of the jet, i.e., the region in which the jet is characterized by a diameter close to that of the inlet nozzle. Beyond this distance, the macroscopic observable $1/<\phi_c>$, namely the inverse time averaged center-line phase field, increases linearly with a slope which, in the dynamical range inspected, depends on the jet Weber number. In particular, we observe an abrupt increase in the angular coefficient of the linear fit as we pass from the first-wind to the second-wind-induced breakup region, i.e., from $We=500$ to $We=2000$. Interesting to note, for $We=2000$ the best fit is obtained by the slope $\sim 1/(D\cdot 6.5)$, close to the value reported in \cite{hussein1994velocity,pope2000turbulent} for momentum jets. This further corroborates the aforementioned observation on the analogy between inertial multicomponent jets and buoyant plumes. 

\begin{figure}
    \centering
    \includegraphics[scale=0.8]{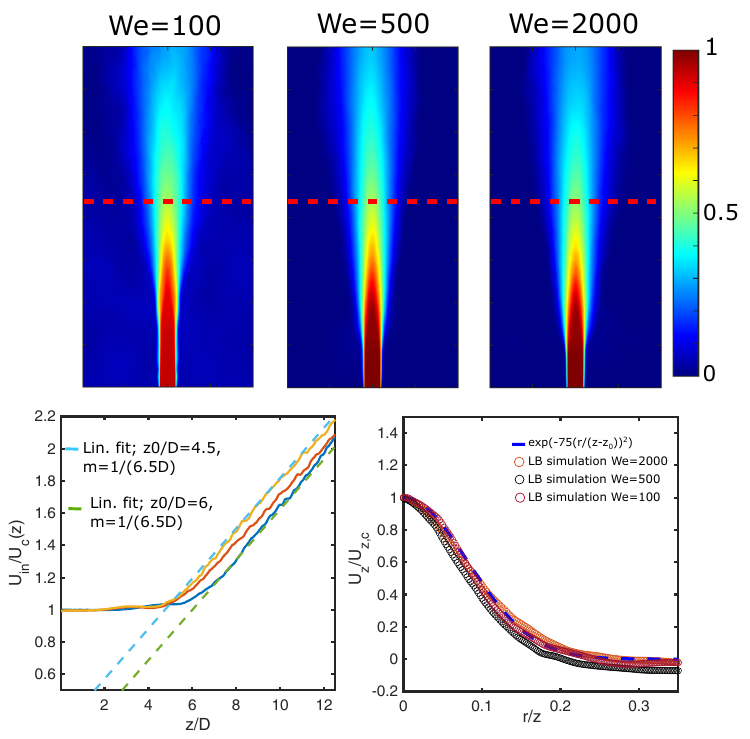}
    \caption{Time averaged velocity fields for increasing Weber numbers. The plots report the inverse of centerline velocity (made non-dimensional by the inlet velocity) versus a non-dimensional distance from the nozzle $z/D$ (left plot) and the mean axial velocity profiles taken at a distance $12D$ from the nozzle (dashed red lines in the velocity fields) (right plot). The profiles fall onto the same curve, denoting the self-similarity of the mean axial velocity in multicomponent turbulent incompressible jets.  }
    \label{fig:self-similar vel}
\end{figure}

It is now instructive to explore whether the presence of an evolving interface between two fluids affects the dynamics of the incompressible jet evolving in a quiescent environment. To this aim, in figure \ref{fig:self-similar vel} the time-averaged velocity fields on a vertical, $x-z$, mid-plane at increasing Weber numbers are reported. In particular, it is interesting to inspect the behavior of the inverse centerline velocity, which, beyond the laminar region, should follow a linear behavior. Indeed, as can be seen in the left plot, the three solid lines (LB simulations) are well approximated by linear fittings featuring the same slope $\sim 1/(6.5 D)$, in agreement with the experimental observation reported in \cite{panchapakesan1993turbulence,boersma1999large} for momentum jets, but slightly different values of the intercepts of the abscissas. From the above observations, we can conclude that the "momentum" potential core appears to be generally decoupled from the interfacial one, following a different dynamics. Such a decoupling is also testified by inspecting the normalized radial velocity profiles taken at a distance of $12D$ downstream of the inlet. As one can see, the mean velocity profiles all fall onto a single Gaussian curve $exp(-75(r/(z-z_0))^2)$, in agreement with the experimental observation reported in \cite{boersma1999large,panchapakesan1993turbulence}.  It is worth mentioning that the observed dynamics are valid for the conditions under which these simulations have been performed (unitary density and viscosity ratios). In conclusion, while the time-averaged phase field shows peculiar features depending on the jet Weber number, the mean velocity profile keeps on exposing a universal behavior that is independent of the competition among surface tension and inertial and viscous drag forces. Future works will investigate the effect of density and viscosity ratios on the above-described time-averaged field quantities.

\section{\label{sec:performances} Performances of the High-order multicomponent thread-safe LB on single GPU }

The multicomponent TSLB implementations have been run on two different accelerators: GeForce RTX 3090 and Nvidia A100 (on Leonardo supercomputer). The former features 24 Gbytes of RAM shared by 10496 Cuda cores with a peak performance $\sim 35$ TeraFLOPS in single precision and a memory bandwidth of $\sim 900$ GB/s, while the latter has 64 Gbytes of RAM shared by 6912 Cuda cores with a peak performance $\sim 19.5$ TeraFLOPS in single precision with a memory bandwidth of  $\sim 2.0 TB/s$. 

The performances of the GPU porting implementations have been measured in billions of lattice update per second (GLUPS), defined as $GLUPS=\frac{n_x n_y n_z n_{steps}}{10^9 T_{sim}}$, where $n_{x,y,z}$ are the number of lattice nodes along the three spatial, dimensions, $n_{steps}$ is the number of simulation time steps and $T_{sim}$ is the run wall-clock time (in seconds) needed to perform the simulation.

The computational domain is composed by $Nx\times Ny \times Nz=128\times 10^6$ grid points.

The code has been ported on GPU by employing  OpenACC directives. The current implementation allows to reach single GPU  performances in line with the state-of-the-art of current LB computing with virtually no extra cost in terms of coding.
More details about the implementation can be found in \cite{MONTESSORIjocs23, montessori2024high}. 

The current simulations deliver  $\sim 1.5$  on RTX3090 and $\sim 2$ GLUPS on the A100 mounted on Leonardo. 
Extension to multi-GPU is currently ongoing and will be the subject of a future publication.

\section{Conclusions}

We presented an efficient thread-safe version of the lattice Boltzmann method aimed at simulating multicomponent systems of immiscible fluids. The model has been employed to simulate incompressible, axisymmetric, two fluids turbulent jets at $Re=5000$ and at increasing Weber, ranging between $We=10$ and $We=2000$, and accurately reproduced breakup regimes over a wide range in the Ohnesorge number, and to give access to highly non-trivial information on the dynamics of high speed, incompressible turbulent jet evolving in quiescent environment.
The multicomponent TSLB GPU implementation presents state-of-the-art performances in lattice Boltzmann computing, thus opening to large-scale simulations of turbulent jet dynamics for aerospace, environmental, and, more broadly, advanced engineering applications. Future development will be aimed at porting the code on multi-GPU via the combination of directive-based programming (OpenACC) and message passing interface (MPI) and in extending the present model to incorporate near-contact interactions to model the effect of surfactant adsorbed at the fluid-fluid interface \cite{montessori2019jfm}.

\section*{Acknowledgements}

A.M. acknowledges the CINECA Computational Grant IsCb3 "recTSLB", IsCb4 "LLBfast" and IsCa9 "3DMPILB" under the ISCRA initiative, for the availability of high performance computing resources and support. M.L. acknowledges the support of the Italian National Group for Mathematical Physics (GNFM-INdAM).


\end{document}